# Direct Optical Probing of Transverse Electric Mode in Graphene

Sergey G. Menabde[1], Daniel R. Mason[1], Evgeny E. Kornev[2], Changhee Lee[2], and Namkyoo Park[1,*]

[1]*Photonic Systems Laboratory, School of ECE, Seoul National University, Seoul 151-744, Korea*

[2]*Organic Semiconductor Laboratory, School of ECE, Seoul National University, Seoul 151-744, Korea*

*nkpark@snu.ac.kr

**Unique electrodynamic response of graphene implies a manifestation of an unusual propagating and localised transverse-electric (TE) mode near the spectral onset of interband transitions. However, excitation and further detection of the TE mode supported by graphene is considered to be a challenge for it is extremely sensitive to excitation environment and phase matching condition adherence. Here for the first time, we experimentally prove an existence of the TE mode by its direct optical probing, demonstrating significant coupling to an incident wave in electrically doped multilayer graphene sheet at room temperature. We believe that proposed technique of careful phase matching and obtained access to graphene's TE excitation would stimulate further studies of this unique phenomenon, and enable its potential employing in various fields of photonics as well as for characterization of graphene.**

Graphene is an atomically thin crystal of carbon atoms with hexagonal lattice[1] which exhibits extraordinary electrodynamic response, and has earned a close attention of optics and photonics communities[2-4]. One of the remarkable properties of graphene is its spectral transition[5] between "metallic" (Drude conductivity) and "dielectric" (broadband and saturable absorbance[6] of ~2.3%) electrodynamic response at energies near $\hbar\omega \sim 2E_F$, where charge carriers' Fermi energy $E_F$ in graphene can be efficiently controlled by doping[5,7-9].

It was predicted[10] that in the spectral transition region, the imaginary part of graphene conductivity ($\text{Im}[\sigma] = \sigma''$) changes sign from positive to negative, having its minimum at $\hbar\omega = 2E_F$ (Fig. 1a). The negative sign of $\sigma''$ has been associated with manifestation of transverse electric[10,11] (TE) propagating electromagnetic mode in both single- and bi-layer[12] graphene, contrary to conventional transverse magnetic (TM), i.e. plasmonic, modes in metals and graphene existing exclusively when $\sigma'' > 0$. Yet, the main difficulty in detecting the TE mode is that its effective index is very close ($\Delta n \sim 0.001$) to that of propagating wave in a bulk material surrounding the graphene, which imposes a necessity of extremely precise phase matching with an excitation wave.

In this article, for the first time eight years after its prediction, we demonstrate direct optical probing of the TE mode supported by graphene, employing modified Otto configuration that allows very precise phase matching of the incident wave to the TE excitation, and electrostatically doped multi-layer graphene stacks at room temperature, performed at the telecommunication wavelength $\lambda_0 = 1.55$ μm.

The physical nature of TE excitations in graphene can be understood as magnetic dipole waves, i.e. self-sustained oscillations of current (Fig. 1b) with no spatial charge density perturbation, while conventional TM plasmons are an electric dipole waves. In analogy to a fundamental guided mode of a high-index dielectric slab waveguide[13] when the slab thickness approaches zero, graphene TE mode is weakly bound but exhibits very low propagation loss[10], and is extremely sensitive to the optical contrast between dielectrics sandwiching the graphene layer[11] which may form the basis of extremely precise sensing of refractive index change in adjacent media[11] with expected sensitivity exceeding $6.7\times10^{-7}$ RIU. It was previously suggested that small effective index of the TE mode (close to that of the surrounding bulk[10]), $n^{eff} = \text{Re}[q]/k_0$ where $q$ is the modal wavenumber and $k_0$ is that in free-space, makes it easily accessible to Otto excitation via evanescent field in attenuated total reflection (ATR) regime[14-16] even at room temperature[10], when doped graphene is sandwiched between two dielectric layers of same refractive index[14].

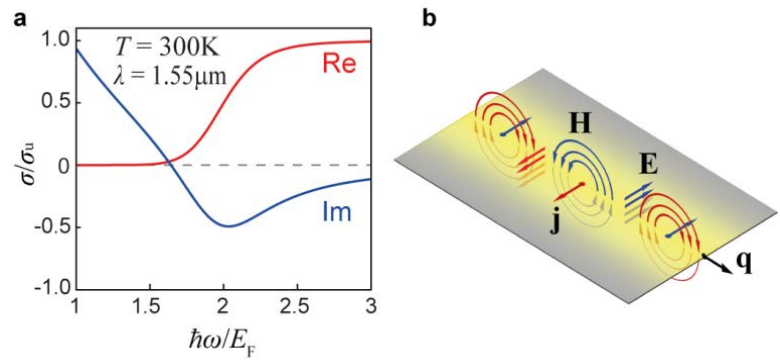

**Figure 1 | Spectral transition of graphene and the TE mode. a**, Graphene sheet conductivity (red – real, blue – imaginary part) in the spectral transition region, obtained using the RPA at indicated parameters. **b**, Schematic representation of graphene TE mode with wavevector **q** as an oscillation of surface current **j** under excitation by transverse electric field **E**.

However, necessity of doping (i.e. special techniques of thereof) makes it difficult to provide symmetry between indices of materials above ($n_2$) and below ($n_3$) the graphene layer. Here we propose a TE mode excitation scheme where graphene is situated between dielectrics with asymmetric indices, $n_2 > n_3$, and placed at distance $d$ from the coupling prism, $n_1$ (Fig. 2a). In such system, in contrary to previously proposed ATR, there available a regime when incident wave experiences total internal reflection (TIR) at the $n_2|n_3$ interface where graphene layer is situated. By employing materials with low optical contrast between $n_2$ and $n_3$, TIR regime with close to 90° angle of incidence for the excitation wave is possible to achieve, which is strongly favourable for adherence of phase matching and efficient TE mode excitation.

**Results**

At first, we consider TE mode behaviour near the cutoff distance $d_{cut}$, first described in our previous work[14], and later confirmed by other research group[16]. The cutoff distance $d = d_{cut}$ denotes the condition of complete TE mode delocalisation in Otto configuration, i.e. when for $n_2 = n_3$ and $d < d_{cut}$, the existence of localised at graphene TE mode is prohibited. When $d > d_{cut}$, the solution of TE mode exists, but starts to be delocalised in the direction of $n_3$ as $d$ approaches to $d_{cut}$. However, increasing of $n_2 > n_3$ makes delocalisation to occur in the opposite direction[14]. Through the balancing of this index asymmetry by a proper selection of distance $d$, a condition of propagating TE mode existence can be preserved.

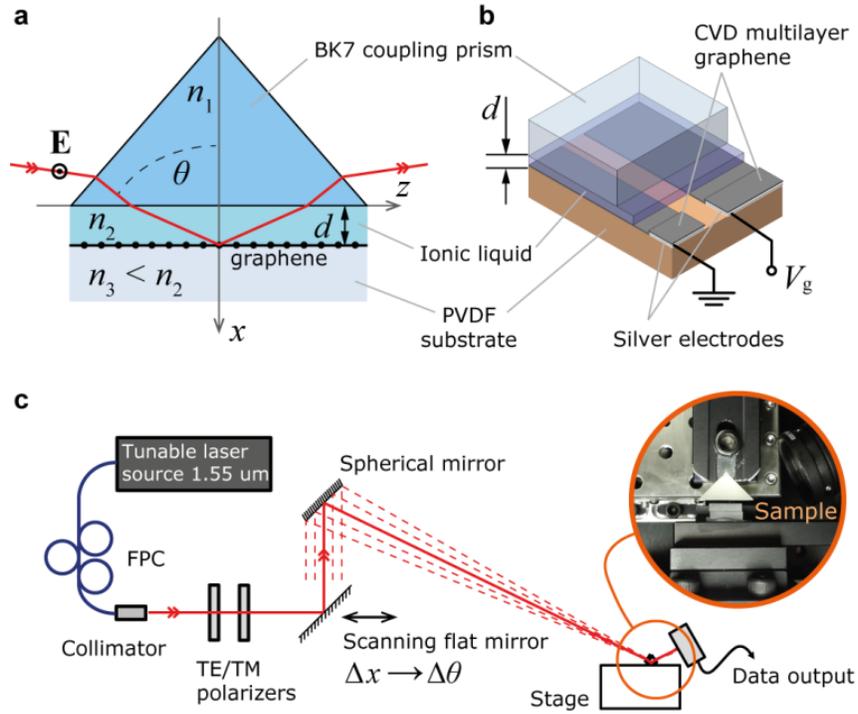

**Figure 2 | Experimental setup. a**, Modified Otto configuration with direct excitation of TE mode in multilayer graphene. **b**, Multilayer graphene sample as inside the experimental setup, where graphene electrodes are electrostatically doped by the gate voltage $V_g$ via the ionic liquid. **c**, Experimental setup employing linear scan transition into the angular scan by a spherical mirror focused at the area where doped graphene is situated. Excitation wave's polarization is controlled by a fibre polarization controller (FPC), and collimated beam diameter is 1 mm, providing an effective angular resolution of the system 0.05 deg.

In a stack of randomly oriented graphene layers (i.e., in a multilayer graphene with rotational faults[17]), an effective optical conductivity of the stack linearly increases proportional to the number of layers[17-20] $N$ ($\sigma^{eff} = N\sigma$). Considering this, we employ a five-layer graphene stack to derive significantly more effective coupling to incident wave[14]. Graphene electrodes are deposited on a polyvinylidene fluoride resin (PVDF; $n_3^{(1.55\mu m)} \approx 1.4045$) with applied silver electrodes (Fig. 2b). Graphene is electrostatically doped by an applied bias voltage via top-gated configuration[21] (essential for high doping values) with an ionic liquid of proper refractive index[22] ($n_2^{(1.55\mu m)} \approx 1.4070$), such that desired index asymmetry condition $n_2 > n_3$ is satisfied. Employing of an ionic liquid instead of an ion gel[23] allows scanning over the distance $d$ between graphene and the coupling prism (BK7; $n_1^{(1.55\mu m)} \approx 1.5006$), approaching desired values of $d \sim \lambda_0$, in order to obtain precise phase matching. To measure an angular reflectance, we employ a setup schematically shown in Fig. 2c, where linear shift of a scanning flat mirror is translated into a high-resolution angular scan by concave spherical mirror of large radius.

Experimentally measured angular reflectance for two different samples is shown in Fig. 3a,b, and is obtained for undoped (dashed lines) and doped (solid lines) 5-layer graphene stack under the bias voltage $V_g$ = 3.8V and 4.5V respectively, for TM (black) and TE (blue) polarized incident wave at $\lambda_0$ = 1.55 μm, and room temperature $T$ = 300 K. While TM curves show no response to doping, TE reflectance of doped graphene exhibits a distinct dip in the vicinity of the critical angle $\theta_c$ and can be observed for both samples, indicating an excitation of a guided TE mode in the structure. It is important to note that in the absence of the graphene, reflectance profile indicates lack of any specific TE mode in the structure under applied experimental conditions, which was confirmed by the control experiment (Fig. 3e).

To confirm the experimental data, and to get an insight into the mode excitation dynamics, we use a numerical model of employed excitation scheme considering angular resolution $\Delta\theta \approx 0.05$ deg (same as experimental setup). We conduct model parameters fitting based on both TE and TM reflectance responses, providing roughly estimated tolerances of $\pm 0.1\lambda_0$ for the $d$, and $\pm 5$meV for the $E_F$ (Fig. 3f). In the model, sheet conductivity of single graphene layer is calculated using random phase approximation[14,24-26] (RPA; see Supplementary Information) in the local limit (i.e. when $q << k_F$; $k_F$ is the Fermi wavenumber), with assumed charge carriers' mobility $\mu = 1 \times 10^4$ cm$^2$(Vs)$^{-1}$. In accordance with the RPA, fitted sheet conductivity is taken as $\sigma_1 = \sigma_u - i0.01\sigma_u$ for the undoped, and $\sigma_2 = 0.3\sigma_u - i0.4\sigma_u$ (model case 1) and $\sigma_2 = 0.7\sigma_u - i0.47\sigma_u$ (model case 2) for the doped graphene, corresponding to doping levels of $E_F \approx 0.42$ eV and 0.38 eV respectively; $\sigma_u = e^2/4\hbar$ is the universal conductivity of graphene[6]. For the 5-layer graphene stack used in the experiment, its effective conductivity is taken as $\sigma^{eff}_{1,2} = 5 \times \sigma_{1,2}$. Additional notes on multilayer graphene doping can be found in Supplementary Information.

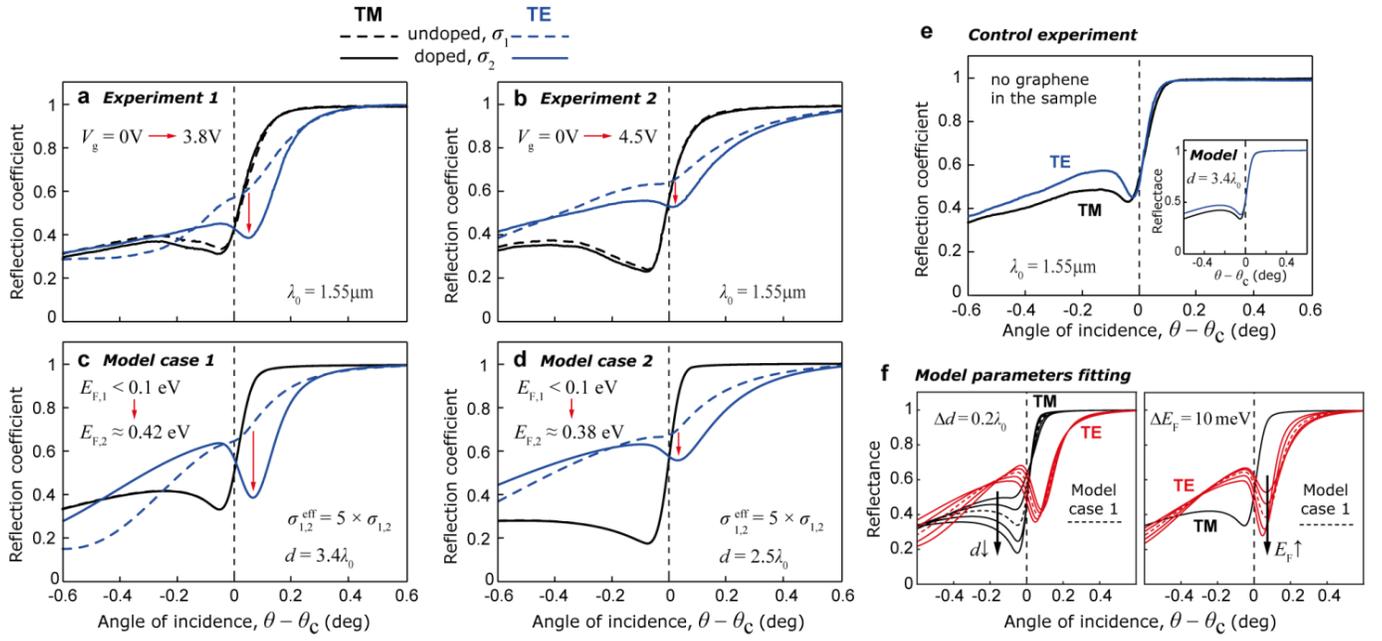

**Figure 3 | Angular reflectance signature of the TE mode. a**, Experimentally measured angular reflectance of sample 1 with 5-layer graphene for TE (blue) and TM (black) incident waves, for undoped (dashed) and doped (solid) graphene under the bias voltage $V_g = 3.8V$; at excitation wavelength 1.55 μ. **b**, Same as in **a**, measured for sample 2, when graphene is doped under the bias voltage $V_g = 4.5V$. **c**, Numerical reflectance data replicating experimental conditions for sample 1 (as in **a**; model case 1), with fitting parameters $d = 3.4\lambda_0$, $n_2 = 1.407$, $n_2 - n_3 = 0.0025$, at indicated values of graphene conductivity; critical angle $\theta_c$ refers to the TIR at $n_2|n_3$ interface. **d**, Numerical reflectance data replicating experimental conditions for sample 2 (as in **b**; model case 2), with similar fitting parameters for $n_{2,3}$ as in **c**, and $d = 2.5\lambda_0$, at indicated values of graphene conductivity. **e**, Angular reflectance measured in the absence of graphene layer under experimental conditions same as in **a**; inset – model numerical reflectance with no graphene for $d = 3.4\lambda_0$. **f**, Numerically demonstrated dependence of the reflectance profile on distance $d$ (left) and graphene doping level (Fermi energy; right) for the basis of model parameters same as in **c** (dashed); arrows indicate change in reflectance profile according to denoted step change in parameters.

Numerical reflectance obtained for the values of $\sigma^{eff}_{1,2}$, and $d = 3.4\lambda_0$ (model case 1) and $d = 2.5\lambda_0$ (model case 2) for both samples is shown in Fig. 3c,d, demonstrating the electrodynamic response in excellent agreement to that experimentally observed. Considering a uniqueness of reflectance profile for given values of $d$ and $\sigma$, along with the estimated tolerances, it can be noted that fitted model parameters are very close to actual experimental conditions, thus confirming the doping of multilayer graphene coming up to negative values of $\sigma''$. Together with the reflectance minima located beyond the critical angle in both experimental and model studies, this indicates an excitation of TE mode in multilayer graphene.

It must be noted that an approximately 1 nm thick Debye layer of ions at the graphene/ionic liquid interface[21] (see Supplementary Information) does not provide detectable interference with the measurements. Considering that the angular reflectance behavior for both TM and TE polarized light perfectly follows theoretical prediction of a model where TE mode is excited in a doped graphene stack (Fig. 3a-d), it is clear that the observed change of the TE reflectance under applied gate voltage is dominantly determined by a change of the optical conductivity in graphene.

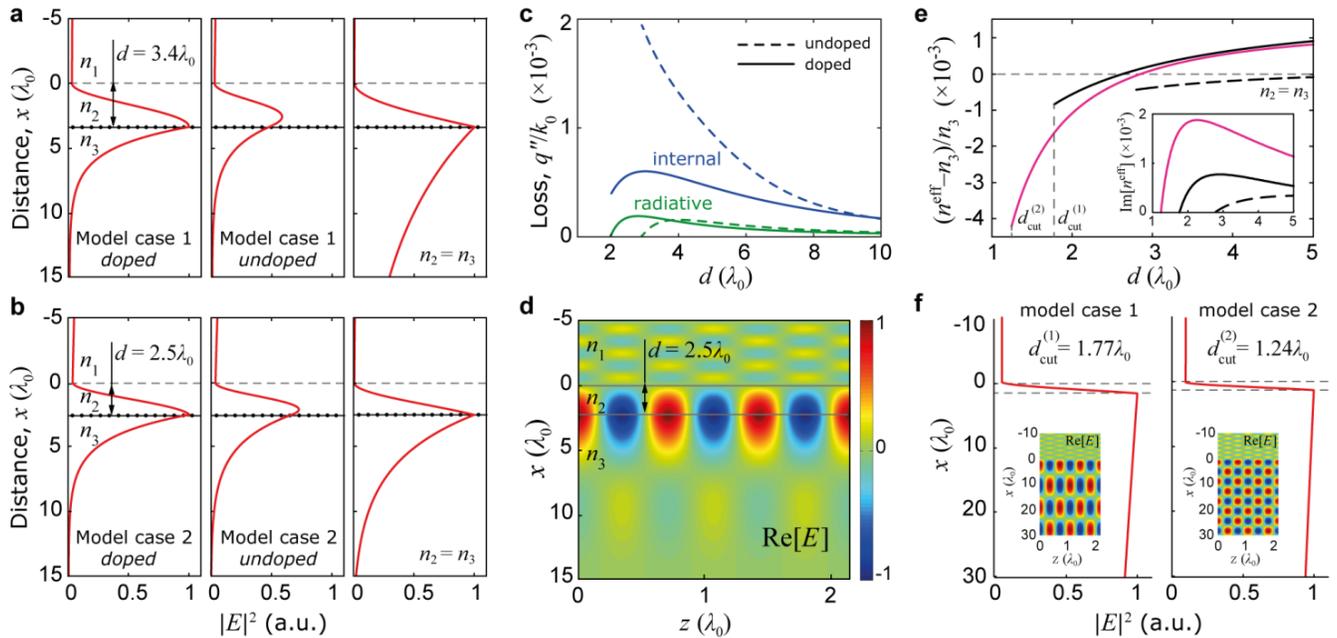

**Figure 4 | Transverse eigenmode of doped graphene. a**. Obtained with a solver, electrical field intensity $|E|^2$ profiles of TE eigenmodes in considered structure; left: TE eigenmode of doped graphene responsible for the experimentally observed TE reflectance dip, for the set of parameters of model case 1; middle: same as on the left, but for undoped graphene – a waveguide mode, with amplitude scaled according to theoretical coupling efficiency 59% of that for doped graphene mode; right: an unperturbed TE mode supported by doped graphene in the considered structure for model case 1 under assumption of $n_2 = n_3$. **b**, Left: same as in **a**, for the set of parameters of model case 2; middle: same as in **a**, with amplitude scaled according to theoretical coupling efficiency 74% of that for doped graphene mode; right: same as in **a**, for the set of parameters of model case 2. **c**, Internal (blue) and radiative (green) losses of the TE eigenmode for the undoped (dashed) and doped (solid) graphene, for the set of parameters of model case 1. **d**, Electric field spatial distribution of the TE eigenmode for the set of parameters of model case 2. **e**. Dispersion of the TE eigenmode for the set of parameters of model case 1 (black), case 2 (pink), and case 1 with $n_2 = n_3$ (dashed) as a function of distance $d$; inset: cumulative losses of the TE eigenmode for all cases. **f**. TE eigenmode's electrical field intensity profile $|E|^2$ shown at the indicated cutoff distances $d = d_{cut}$, for model cases 1 and 2; insets: electrical field spatial distribution demonstrated for each cutoff case.

Further numerical investigation with a solver reveals a consistent presence of the TE eigenmode supported irrespectively of the structure variations, proving its existence solely due to the presence of multilayer graphene stack. Fig. 4a,b demonstrate an electric field intensity $|E|^2$ profile of this mode. Leftmost profiles correspond to reconstructed experimental conditions (Fig. 3c,d), and demonstrate the profile of the TE mode responsible for the observed reflectance dip at doped multilayer graphene (Fig. 3a,b). This doped graphene mode, although perturbed by the leaky term, is confined to graphene plane in contrary to the waveguide mode supported with presence of undoped lossy graphene in an asymmetric index (the midst in Fig. 4a,b). An unperturbed TE graphene mode is also supported by the structure with considered parameters, but only when $n_2 = n_3$, thus being tricky to be excited; its profile is the rightmost in Fig. 4a,b.

Graphene sheets in a multilayer stack are separated by approximately 0.3 nm distance[20]. Considering huge (about three orders of magnitude) difference between the size of the TE mode (Fig. 4a,b) and the actual thickness of the 5-layer graphene stack, experimentally detected TE mode is supported by a thin (i.e., with effectively zero thickness) layer of graphene with effective conductivity $\sigma^{eff}$ proportional to the number of layers[14,17-20]. At the same time, the physical nature of the mode is that of graphene TE mode. Coupling between TE modes of two separated graphene layers is explicitly discussed in the authors' previous work[14].

Figure 4c displays the effect of doping on losses associated with the TE mode as a function of distance $d$. Due to reduction of internal losses (blue) in doped graphene (solid lines) at shorter $d$, their magnitude becomes comparable with that of radiative losses (green), providing significantly enhanced coupling to incident light[13,14,27], and producing experimentally observed distinct reflectance dip. As one can note, cumulative losses of the excited mode are still very low, though higher than in case of $n_2 = n_3$ (see inset in Fig. 4e), giving its propagation length $L = \lambda_P/(2\pi q'') \sim 1000$ of plasmonic wavelengths.

TE mode's electric field spatial distribution for the model case 2 is demonstrated at Fig. 4d, showing significant asymmetry along the $x$-$z$ axes, revealing mode expansion by several $\lambda_0$ into the bulk on both sides. Obtained with the solver dispersion of the TE mode for model case 1 (black), case 2 (pink), and case 1 with $n_2 = n_3$ (dashed) is demonstrated in Fig. 4e, with losses shown in the inset. Observed in Fig. 4e for model cases 1 and 2, the dispersion cutoff is associated with complete delocalization (leaking) of modes in the direction of $n_3$, following prediction for the symmetrical ($n_2 = n_3$) configuration[14]. Modes' delocalization is demonstrated in Fig. 4f where field's intensity and spatial distribution are shown for distances $d \approx d_{cut}$ for considered model cases. It is also clearly demonstrated (Fig. 4e) that cutoff distance in case when $n_2 = n_3$ is significantly larger comparing to that in employed experimental configuration, preventing efficient coupling to the TE mode.

To summarize, we conclude that exotic TE mode in graphene can be successfully excited in multilayer graphene at room temperature. Its first experimental observation is achieved employing 5-layer doped graphene in modified Otto configuration with very precise phase matching capability. Besides, we demonstrate successful doping of multilayer graphene up to significant values of $E_F$ using an ionic liquid, confirming predicted earlier increase of effective graphene conductivity[17-20]. We also believe that optical sensing of TE excitation in doped graphene can be used as a handy technique for characterisation of doping in multilayer graphene, and attract more attention to this unique phenomenon in the essential spectral transition domain.

**Methods**

**Sample manufacturing and measurements** In this study we use PVDF film with thickness of 0.254 mm was purchased from CS Hyde; silver electrodes were deposited manually using Pelco® colloidal silver liquid manufactured by Ted Pella, Inc. (product No. 16031), and cured at room temperature during 24 hours. 12.5×12.5 mm² area five layer graphene was manufactured by chemical vapour deposition[28] (CVD) and deposited layer by layer in acetone, and was purchased from Graphene Square, Inc. The ionic liquid 1-ethyl-3-methylimidazolium bis(trifluoromethylsulfonyl)imide was used as obtained from manufacturer, purchased from Sigma-Aldrich.

Infrared reflectance measurements were performed with a tunable diode laser at 1.55 µm wavelength and a power meter head with removed cover glass to avoid interference effects under varying angle. Gold-covered spherical mirror of 1250 mm radius was mounted at fixed stage while angular scan was conducted by a flat mirror on a moving linear stage (Fig. 2c), providing, together with collimated laser beam with diameter of 1 mm, an effective angular resolution of 0.05 deg, while mechanically limited angular resolution allowed by the setup is $3\times10^{-5}$ deg. All measurements were performed under normal ambient conditions at 300 K.

**Analytical model** In the numerical model, the excitation scheme shown in Fig. 2(a) is considered as a three-layered structure with an effectively semi-infinite top (coupling prism, $n_1$) and bottom (PVDF substrate, $n_3$) layers, where the plane $x = 0$ corresponds to the $n_1|n_2$ interface, and the multilayer graphene stack of effectively zero thickness occupies the $x = d$ plane. TE-polarized plane wave is incident onto the $n_1|n_2$ interface at angle $\theta$. The *ansatz* is made that the electric field in each medium $m = 1,2,3$ takes the form $\mathbf{E}^{(m)} = \hat{\mathbf{y}} E_m(x) \exp(iqz - i\omega t)$, where,

$$E_1(x) = A_1 \exp(ik_1 x) + B_1 \exp(-ik_1 x), x < 0;$$
$$E_2(x) = A_2 \exp(ik_2[x-d]) + B_2 \exp(-ik_2[x-d]), 0 < x < d;$$
$$E_3(x) = A_3 \exp(ik_3[x-d]) + B_3 \exp(-ik_3[x-d]), x > d.$$

Substituting the *ansatz* into the Helmholtz equation $\nabla^2 \mathbf{E}^{(m)} + k_0^2 \mathbf{E}^{(m)} = 0$ (where $k_0 = \omega/c$) we obtain $k_m = \sqrt{k_0^2 n_m^2 - q^2}$, where $q = n_1 k_0 \sin(\theta)$ is the projection of the incident wavevector onto the *z*-axis. Then, assuming the harmonic time dependence $\exp(-i\omega t)$ of the magnetic field **H**, substituting it into the Maxwell curl equation $\nabla \times \mathbf{E} + \mu_0 \partial \mathbf{H}/\partial t = 0$, we apply electromagnetic boundary conditions $\hat{\mathbf{x}} \times (\mathbf{E}^{(m+1)} - \mathbf{E}^{(m)}) = 0$, $\hat{\mathbf{x}} \times (\mathbf{H}^{(m+1)} - \mathbf{H}^{(m)}) = \mathbf{K}$ at the $n_1|n_2$ and $n_2|n_3$ interfaces, noting that $\mathbf{K}(x = 0) = 0$ and $\mathbf{K}(x = d) = \sigma^{\text{eff}} \mathbf{E}(x = d)$, thus obtaining the system of linear equations in complex amplitudes $A_m$ and $B_m$. Taking $B_3 = 0$ (i.e., absence of a reflected wave in medium 3) and $A_1 = 1$ (i.e., continuous incident wave), solving the system, we obtain the reflection coefficient of the incident wave $R = |r|^2$ ($r = B_1/A_1$). The same system of linear equations in complex amplitudes is used in a solver to find eigenmodes solutions supported by the structure, solving the determined system in the absence of the incident wave $A_1 = 0$.

**Acknowledgements**

This work was supported by the National Research Foundation of Korea (NRF) under the Ministry of Science, ICT & Future Planning; the Global Frontier Program NRF-2014M3A6B3063708, and the Global Research Laboratory (GRL) Program K20815000003 (2008-00580), all funded by the South Korean government.

**Author contributions**

S. M. developed numerical model, proposed and performed experimental work. D. M. performed theoretical and numerical calculations, E. K. and C. L. assisted in samples manufacturing, and N. P. supervised the work, participated in discussion and in writing the manuscript.